\documentclass[prl,aps, superscriptaddress, twocolumn,showpacs,preprintnumbers,amsmath,amssymb, 10pt]{revtex4}
\usepackage{epsfig,amsmath}
\usepackage{amssymb,amsmath,wasysym}

\usepackage[english]{babel}
\usepackage{graphicx}

\begin{document}
%opening
\title{Non-spectral relaxation in one dimensional Ornstein-Uhlenbeck processes}
\author{R. Toenjes}
\email{toenjes@uni-potsdam.de}
\affiliation{Institute of Physics and Astronomy, Potsdam University, 14476 Potsdam-Golm, Germany}
\author{I.M. Sokolov}
%\email{igor.sokolov@physik.hu-berlin.de}
\affiliation{Institut f\"ur Physik, Humboldt Universit\"at zu Berlin, 12489 Berlin, Germany}
\author{E.B. Postnikov}
%\email{postnicov@gmail.com}
\affiliation{Department of Theoretical Physics, Kursk State University, 305000, Kursk, Russia}

\begin{abstract}
The relaxation of a dissipative system to its equilibrium state often shows a multiexponential 
pattern with relaxation rates, which are typically considered to be independent of the initial condition.
The rates follow from the spectrum of a Hermitian operator obtained by a similarity transformation 
of the initial Fokker-Planck operator. 
However, some initial conditions are mapped by this similarity transformation to functions
which grow at infinity. These cannot be expanded in terms of the eigenfunctions of an Hermitian operator, 
and show different relaxation patterns. 
Considering the exactly solvable examples of Gaussian and generalized L\'evy Ornstein-Uhlenbeck processes (OUPs)
we show that the relaxation rates belong to the Hermitian spectrum 
only if the initial condition belongs to the domain of attraction of the 
stable distribution defining the noise. 
While for an ordinary OUP initial conditions leading to non-spectral relaxation can be considered exotic, for generalized OUPs driven by L\'evy noise these are the rule.
\end{abstract}


\maketitle

The relaxation of a physical system, prepared in a non-equilibrium state, to the equilibrium often shows a multiexponential 
pattern with decrements of single exponentials defining the relaxation rates. These rates are usually considered an intrinsic property of the system, independent of initial 
conditions and they follow from the spectrum of a Hermitian Hamiltonian operator obtained by a similarity transformation of the Fokker-Planck (FP) operator governing the evolution of the probability density. 
Methods of spectral analysis are central in physics, in particular in quantum mechanics and in the theory of oscillations, and are universally employed to the solution of linear problems.
Thus the discussion of the spectrum of the FP operator is often the first step in the solution of the FP equation and the investigation of its relaxation properties \cite{Risken, Brockett, Felderhof}.
\\ %\\
As we proceed to show, this first step might not deliver a complete picture.
Initial distributions which are not mapped to square integrable functions by the similarity transformation, cannot be expanded in terms of the eigenfunctions of the corresponding Hamiltonian operator and will 
therefore relax at rates that may not be given by the Hermitian spectrum. It is in this sense that we use the term non-spectral relaxation.
The smallest non-spectral rate can be smaller than the smallest spectral relaxation rate
and thus dominate the relaxation behavior over the whole time range. 
\\ %\\
Although the effect of non-spectral relaxation can be observed under quite general conditions, in this Letter we concentrate on simplest, exactly solvable examples of Ornstein-Uhlenbeck processes (OUPs) describing 
the coordinate of an overdamped particle in a harmonic potential driven by a white noise. Because the OUP generally approximates random processes in the vicinity of a stable stationary point it is a very important 
analytical tool in many fields of research, from statistical physics \cite{Felderhof, Coffey, Gillespie}, to theoretical neuroscience \cite{NeuroOU}, ecology \cite{ButlerKing} and economics \cite{BNielsen}. 
Since the assumption of Gaussian fluctuations is violated in many non-equilibrium systems, in recent years the attention has been shifted to non-Gaussian statistics arising from L\'evy noise  \cite{BNielsen,Dybiec}. In the present paper we consider both,
the standard OUP driven by a Gaussian noise, and a generalized OUP driven by symmetric, white L\'evy noise. 
Even for the Gaussian OUP, the non-spectral relaxation of broad initial distributions has, 
surprisingly, never been considered in the textbooks and the applied mathematical literature \cite{Risken, Brockett, Felderhof}.
For the L\'evy OUP, the similarity transformation to the quantum harmonic oscillator Hamiltonian, which we present here,
allows us to define a Hermitian spectrum for the corresponding fractional FP operator and hence to distinguish between spectral and non-spectral relaxation in this generalized case.
The effect of non-spectral relaxation found in this letter is of different nature from the non-spectral relaxation in the presence of multiplicative noise described in \cite{MultiNoise} and from slow, 
non-exponential relaxation specific to subdiffusive processes \cite{Barakai}.
\\ %\\
The time dependent probability density $p(x,t)$ for a Gaussian diffusion process in a one-dimensional potential $U(x)$ solves a FP equation \cite{Risken} of the form
\begin{equation}	\label{FPE}
	\frac{\partial}{\partial t} p(x,t) = \frac{\partial}{\partial x}\left[ U'(x) p(x,t) \right] + \frac{\partial^2}{\partial x^2} p(x,t) = \hat{L} p(x,t),
\end{equation}
where the time has been scaled to units of the inverse diffusion constant. 
The time independent, stationary solution is given as $p_{st}(x)=\frac{1}{Z}e^{-U(x)}$ where $Z$ is determined by normalization. The time evolution of the transformed function
\begin{equation}	\label{Eq:HTransform}
	\psi(x,t) = p_{st}(x)^{-\frac{1}{2}}p(x,t) = \frac{1}{\sqrt{Z}} e^{\frac{1}{2}U(x)} p(x,t)
\end{equation}
is given by a Hermitian Operator $\hat{H}$
\begin{equation}
	-\frac{\partial}{\partial t} \psi(x,t) = \hat{H} \psi(x,t) = \left[V(x) - \frac{\partial^2}{\partial x^2}\right] \psi(x)
\label{Schroed}
\end{equation}
with $V(x)=[\frac{1}{4}U'(x)^2-\frac{1}{2}U''(x)]$ \cite{Risken}. Equation (\ref{Eq:HTransform}) defines a similarity transformation between $\hat{L}$ and $-\hat{H}$ and, hence, a transformation $\psi = \hat{S} p$ 
from the space of the solutions of the FP equation to the space of the solutions of the Schroedinger-like equation (\ref{Schroed}).
The eigenvalues $-\lambda$ of the Hamiltonian $\hat{H}$ are real valued and the corresponding eigenfunctions 
$\psi_\lambda(x)$ form a basis in the Hilbert space of square integrable functions. A distribution $p(x,t)$ that can be expanded into the transformed eigenfunctions 
$\varphi_\lambda(x)=\psi_\lambda(x)\sqrt{p_{st}(x)}$ will relax at rates that are given by the eigenvalue spectrum of $\hat{H}$. We call this a {\it spectral} relaxation pattern. 
However, from Eq.(\ref{Eq:HTransform}) follows that only those distributions $p(x,t)$ transform into a square integrable function $\psi(x,t)$ which decay sufficiently faster at infinity than $1/\sqrt{p_{st}(x)}$ grows \cite{Brockett}. 
If $U(x)$ goes faster to infinity than logarithmically, $p(x,t)$ must decay exponentially. In this case, the existence of all moments is a necessary and, indeed, sufficient condition for spectral relaxation. 
Other, fully legitimate probability density functions, for instance a Cauchy distribution $p(x,t_0) = (1/\pi) (x^2+1)^{-1}$, as initial distribution for the FP equation, 
are mapped to functions that grow rapidly at infinity and cannot be expanded into the square integrable eigenfunctions of the Hamiltonian. The relaxation for such initial
conditions does not have to be spectral. Since one is usually interested in the Green's function of the system, which is the conditional probability density $p(x,t+\tau|x_0,t)$, 
square integrability of the transformed initial condition is always assumed and other cases have never been considered, because they seem exotic or even unphysical.
For the OUP with a linear restoring force, given by a mobility coefficient $\nu$, the potentials are $U(x)=\frac{1}{2}\nu x^2$ and $V(x)=\frac{1}{4}\nu^2x^2-\frac{1}{2}\nu$. 
The spectral relaxation rates are given by the energy eigenvalues $-\lambda_n=n\nu$, $n\in\mathbb{N}$ of the quantum harmonic oscillator with ground state energy zero.
\\ %\\
Let us proceed to show that the fractional FP operator of the L\'evy OUP \cite{Metzler1999, Jespersen1999} can also be mapped to the Hamiltonian of the quantum harmonic oscillator. 
The FP equation for the probability distribution of a Levy flight in a harmonic potential reads
\begin{equation}
 \frac{\partial p}{\partial t} = \frac{\partial}{\partial x} \left[ \nu x p(x,t) \right] + \Delta^{\mu/2} p(x,t) = \hat{L}_\nu^\mu p(x,t)  ,
 \label{GFPE}
\end{equation}
with the parameter $0 < \mu \le 2$ in the fractional derivative corresponding to the index of the stable law defining the L\'evy noise. The fractional Laplacian is defined by its action in Fourier space: $\Delta^{\mu/2}p(x) \to - |k|^\mu p(k)$, where it is diagonal. 
We write $\hat{L}_\nu^\mu$ for the corresponding fractional FP operator depending on the noise parameter $\mu$ and the mobility $\nu$.
Eq.(\ref{FPE}) with $U(x)=\frac{1}{2}\nu x^2$ is a special case of Eq.(\ref{GFPE}) with $\mu=2$. 
%The relaxation rates are eigenvalues of the operator $\hat{L}_\nu^\mu$. 
In Fourier space Eq.(\ref{GFPE}) is an evolution equation for the characteristic function $p(k,t)=\textrm{E}_p[e^{ikx}]$. There it has the form
\begin{equation} 
 \frac{\partial}{\partial t}p(k,t) = -\nu k\frac{\partial}{\partial k}p(k,t)- |k|^{\mu}p(k,t).
\label{GFPEk}
\end{equation}
By simply rescaling the argument with a diagonal transformation $\hat{T}_\alpha$ with integral kernel \mbox{$T_\alpha(\kappa,k)=\delta(|\kappa|^{1/\alpha}\textrm{sign}(\kappa)-k)$},
\begin{equation}
	\left[\hat{T}_\alpha p\right](\kappa,t) = \int T_\alpha(\kappa,k)p(k,t) dk = p(|\kappa|^\frac{1}{\alpha}\textrm{sign}(\kappa),t),
\end{equation}
and using the chain rule $k\partial_k [\hat{T}_\alpha p] = \alpha\kappa \partial_\kappa [\hat{T}_\alpha p]$ we find that, with $\alpha=\mu/2$, the transformed functions follow a non-fractional FP equation
\begin{equation}
	\frac{d}{dt}[\hat{T}_{\frac{\mu}{2}} p] = -\nu\frac{\mu}{2} \kappa \frac{\partial}{\partial\kappa} [\hat{T}_{\frac{\mu}{2}} p] + \frac{\partial^2}{\partial\kappa^2} [\hat{T}_{\frac{\mu}{2}} p] = \hat{L}^2_{\nu\frac{\mu}{2}} [\hat{T}_{\frac{\mu}{2}} p].
\end{equation}
For any $\alpha>0$ the transformation is defined everywhere, it preserves the value $p(k=0)=[\hat{T}_\alpha p](\kappa=0)$, i.e., the normalization in coordinate space, 
and it has $\hat{T}_\alpha^{-1} = \hat{T}_{1/\alpha}$ as the inverse. Indeed, it defines the similarity transformation $\hat{T}_{\mu/2} \hat{L}_\nu^\mu \hat{T}_{\mu/2}^{-1} = \hat{L}^2_{\nu \mu/2}$ 
of the fractional FP operator $\hat{L}_\nu^\mu$ to that of the non-fractional OUP $\hat{L}^2_{\nu \mu/2}$ with the coefficient of restoring force depending on the noise parameter $\mu$. 
In coordinate space, the transformation $\hat{T}_\alpha$ is an integral transform with the kernel
\begin{equation}
T_{\alpha}(\chi,x)=\frac{1}{2\pi}\int_{-\infty}^{\infty}e^{i\kappa \chi - i|\kappa|^{\frac{1}{\alpha}}\mathrm{sign}(\kappa) x}d\kappa.
\label{direct_transform}
\end{equation}
Thus, applying the transformations $\hat{S}$ and $\hat{T}_{\mu/2}$ in sequence, one can transform the fractional operator $\hat{L}_\nu^\mu$ to the Hamiltonian of the quantum harmonic 
oscillator with the harmonic eigenvalue spectrum $-\lambda_n=n\nu\frac{\mu}{2}$.
\\ %\\
The spectrum of a Hermitian operator and its eigenfunctions are determined by the properties of the Hilbert space it is operating on. In the case of square integrable functions, there are selection 
rules that constrict the possible eigenvalues, and the corresponding eigenfunctions form a complete basis. Given the similarity transformation between the FP operators for the one 
dimensional Gaussian diffusion process in any potential or the generalized L\'evy OUP and a quantum mechanical Hamiltonian, it is tempting to use the same selection rules in 
order to also resolve the identity in the space $L^1$ of integrable solutions of the initial FP equation. But one has to be aware that this is only possible in a subspace of $L^1$. 
However, both, the solution of the eigenvalue problem, and the complete time dependent solution of the fractional Fokker-Planck equation for the L\'evy OUP, 
can be found analytically. We therefore use this analytically tractable example as a showcase for non-spectral relaxation, which could otherwise not be explained by Hermitian spectral theory.
\\ %\\
Since we only consider real valued functions in coordinate space, we can restrict the eigenfunctions $\varphi_\lambda(k)$ of the FP operator $\hat{L}^\mu_\nu$ in Fourier space 
to those for which $\varphi_\lambda(-k)=\varphi_\lambda(k)^*$ holds. The eigenvalue problem $\hat{L}^\mu_\nu \varphi_\lambda = \lambda\varphi_\lambda$ is solved via separation of variables by 
any $\lambda\in\mathbb{R}$, $\lambda\le 0$ and $a_\lambda,b_\lambda\in\mathbb{R}$ as
\begin{equation}
	\varphi_\lambda(k) = \left[a_\lambda + i b_\lambda \textrm{sign}(k)\right] |k|^{-\frac{\lambda}{\nu}} e^{-\frac{1}{\nu\mu}|k|^\mu}.
\end{equation}
A nonzero coefficient $a_\lambda$ means that the eigenfunction in coordinate space has a nonzero even part and a nonzero 
$b_\lambda$ contributes to the odd part of $\varphi_\lambda(x)$. For the symmetric L\'evy flight in a symmetric potential, the stationary solution must be even, and we find 
$p_{st}(k)=\varphi_0(k)=\exp(-\frac{1}{\nu\mu}|k|^\mu)$ and thus
\begin{equation}	\label{Eq:EigenSolve}
	\varphi_\lambda(k) = \left[a_\lambda + i b_\lambda \textrm{sign}(k)\right] |k|^{-\frac{\lambda}{\nu}} p_{st}(k).
\end{equation}
The characteristic function $p(k,t+\tau)$ at time $t+\tau$ is the unique solution of the fractional FP equation (\ref{GFPEk}) 
with given initial characteristic function $p(k,t)=p_0(k)$ at time $t$. It is found by the method of characteristics and yields
\begin{equation} \label{Eq:CharacterSolve}
	p(k,t+\tau) = \frac{p_0(ke^{-\nu\tau})}{p_{st}(ke^{-\nu\tau})} p_{st}(k).
\end{equation}
Comparing Eqs.(\ref{Eq:EigenSolve}) and (\ref{Eq:CharacterSolve}) we see, that $p(k,t+\tau)$ has an expansion into eigenfunctions of $\hat{L}_\nu^\mu$ if the ratio $p_0/p_{st}$ can be expanded as
\begin{equation} \label{Eq:p0pstfrac}
	\frac{p_0(ke^{-\nu\tau})}{p_{st}(ke^{-\nu\tau})} = \sum_{\lambda} \left[a_\lambda + ib_\lambda\textrm{sign}(k)\right] |k|^{-\frac{\lambda}{\nu}} e^{\lambda\tau}.
\end{equation}
In more general cases, the sum may be replaced by an integral with respect to an appropriate measure over the non-positive real numbers. 
Note, that both, the initial distribution and the stationary distribution determine the relaxation rates to the equilibrium. 
Two-point correlation functions of observables with finite expectation and variance at equilibrium require the conditional probability distribution $p(x,t+\tau|x_0,t)$ which is the solution of 
the FP equation with a delta distribution $p_0(x)=\delta(x-x_0)$ at an initial time $t$. The asymptotic relaxation rate of these correlation 
functions can be different from the relaxation rates of probability densities with other initial distributions. 
\\ %\\
It is beyond the scope of this letter to study the 
conditions under which such an expansion exists. Instead, we show an example for which the expansion into an absolutely convergent series is known, and 
where the contributing eigenvalues $\lambda$ are not identical with the harmonic spectrum $\lambda_n=-n\nu\frac{\mu}{2}$, $n\in \mathbb{N}$. 
Let us consider a L\'evy stable distribution of index $\alpha$ centered around a point $x_0$, which has the characteristic function $p_0(k)=\exp\left(ikx_0 - \sigma_0|k|^{\alpha}\right)$. 
The fraction (\ref{Eq:p0pstfrac}) has the absolutely convergent series expansion
\begin{equation}
	\frac{p_0(ke^{-\nu\tau})}{p_{st}(ke^{-\nu\tau})} = \sum_{l,m,n=0}^\infty c_{lmn} |k|^{-\frac{\lambda_{lmn}}{\nu}}e^{\lambda_{lmn}\tau}
\end{equation}
with
\begin{equation}
	c_{lmn}=\frac{1}{l!m!n!}\left(ix_0\textrm{sign}(k)\right)^l\left(-\sigma_0\right)^m\left(\frac{1}{\nu\mu}\right)^n
\end{equation}
and relaxation rates
\begin{equation}
	\lambda_{lmn} =-\nu(l+m\alpha+n\mu), \qquad l,m,n\in\mathbb{N}. 
\end{equation}
Note, that odd eigenfunctions occur in the expansion only for odd $l$ and asymmetric initial conditions $x_0\ne 0$, 
and the smallest eigenvalue of an odd eigenfunction is simply given by the deterministic exponential relaxation of the mean to its stationary 
value zero at rate $\nu$, independently of the noise parameter $\mu$. Non-spectral relaxation rates are observed, whenever the initial distribution does not belong to the domain of attraction of the stationary 
distribution (as a stable law), i.e. for $\alpha\ne\mu$, or $\mu<2$ and $x_0\ne 0$. A delta distribution at the origin, i.e. $x_0=0$ and $\sigma_0=0$, 
can be expanded into the even eigenfunctions corresponding to the harmonic eigenvalues $\lambda_{2n}=-n\nu\mu$, because this special case belongs to the domain 
of attraction of all stable laws. The corresponding expansion was found in \cite{Jespersen1999}, which could mistakenly be interpreted as a hint that the complete eigenvalue spectrum of the FP operator for the L\'evy OUP is, in fact, harmonic.
\\ %\\
Given the time dependent solution (\ref{Eq:CharacterSolve}) of the FP equation, in order to demonstrate non-spectral relaxation, one can look at the
relaxation of the expected values of appropriate observables to their stationary values. Instead, here we use the $L^2$-distances
\begin{eqnarray} \label{Eq:CM}
 \Delta^2_{+} &=& \int_{-\infty}^{\infty} (p^+(x,\tau) - p_{st}(x))^2 dx \\
&=& \frac{1}{2\pi} \int_{-\infty}^{\infty} (p^+(k,\tau) - p_{st}(k))^2 dk \label{Distanceeven}, \nonumber \\ 
 \Delta^2_{-} &=& \int_{-\infty}^{\infty} p^-(x,\tau)^2 dx = \frac{1}{2\pi} \int_{-\infty}^{\infty} |p^-(k,\tau)|^2 dk \nonumber \label{Distanceodd},
\end{eqnarray}
which are the square norms of the difference between the even and the odd parts of the time dependent probability densities $p(x,\tau)=p^+(x,\tau)+p^-(x,\tau)$ and the corresponding parts of the stationary density 
$p_{st}(x)=p_{st}^+(x)$, which has no odd part. 
\begin{figure}[!t] 
\includegraphics[width=7cm]{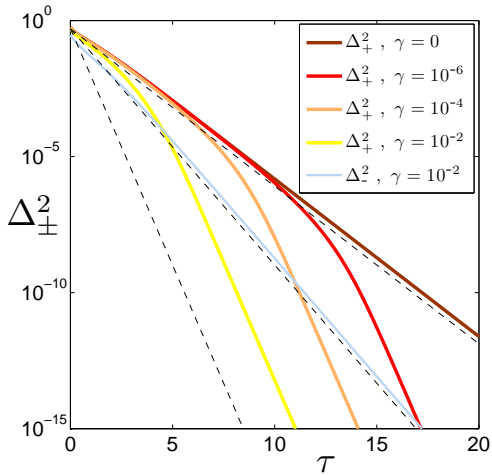}
\caption{(color online) Semilogarithmic plot of the $L^2$-distances $\Delta^2_+$ and $\Delta^2_-$ between, respectively, the even and odd parts of the 
time-dependent and the stationary probability density, in case of the OUP with Gaussian white noise ($\nu=1$, $\mu=2$) and shifted, 
tempered $\alpha$-stable initial distribution (Eq.(\ref{Eq:TemperShift}), $\sigma_0=1$, $\alpha=2/3$, $x_0=1$). The asymptotic relaxation rate of the square distance 
of the odd part (thin solid, blue line) is $-2\lambda_1=2\nu=2$, independent of the cut-off parameter $\gamma$. The square distance of the even part displays 
a cross-over from slow, non-spectral decay at a rate $2\alpha=4/3$ to spectral relaxation at rate $-2\lambda_2=2\mu=4$. The transient is longer for smaller values 
of $\gamma$, i.e. broader distributions. Here we have plotted $\Delta^2_+$ for $\gamma=0$, $10^{-6}$,$10^{-4}$ and $10^{-2}$ (bold, solid lines). 
Using a L\'evy stable distribution as initial condition, i.e. $\gamma=0$, non-spectral relaxation of $\Delta^2_+$ is observed at all times. 
The dashed lines are exponential functions $\frac{1}{2}\exp(\lambda\tau)$ with $\lambda=-4$, $-2$ and $-4/3$, drawn for comparison. 
The $L^2$-distances where calculated according to Eqs.(\ref{Eq:CharacterSolve}), (\ref{Eq:CM}) and (\ref{Eq:TemperShift}) by numerical quadrature in Fourier space.}
\label{Fig1:Delta}
\end{figure}
The relaxation rates of these $L^2$-distances assume twice the value of the eigenvalues in the expansion (\ref{Eq:p0pstfrac}) of $p_0/p_{st}$.
As a particularly striking example, in Fig.\ref{Fig1:Delta}, we plot the $L^2$-distances for the case of the Gaussian OUP 
($\mu=2$) with a smoothly tempered, L\'evy stable initial distribution \cite{SteMa,Koponen} shifted to a point $x_0\ne 0$. The characteristic function for the jump size 
distribution of the truncated L\'evy flight with exponential cutoff was found in \cite{Koponen} and for $\alpha\ne 1$ is given by
\begin{equation}	\label{Eq:TemperShift}
	p_0(k) = \exp\left(ikx_0 - \sigma_0 \frac{\textrm{Re}\left[\left(\gamma + i|k|\right)^\alpha\right]-\gamma^\alpha}{\cos\left(\alpha\frac{\pi}{2}\right)}\right).
\end{equation}
Since for $\gamma\ne 0$ all derivatives at $k=0$, and hence, all moments exist, we expect spectral relaxation at the asymptotic rates $-2\lambda_1=2\nu$ for the odd $L^2$-distance $\Delta^2_{-}$, and $-2\lambda_2=4\nu$ for the even $L^2$-distance $\Delta^2_{+}$. In Fig.\ref{Fig1:Delta} we 
observe that $\Delta^{2}_-$ relaxes at the spectral rate $-2\lambda_1=2\nu$ independently of $\gamma$. 
On the other hand, the decay of $\Delta^{2}_+$, 
which is expected to be faster than its odd counterpart, is delayed during a transient that depends on $\gamma$, and may be considerably slower than $4\nu$, 
depending on the index $\alpha$ of the initial distribution. In fact, the transient decay rate is given by twice the smallest eigenvalue $\lambda=-\alpha$ 
used in the expansion of the $\alpha$-stable law approximated by the initial distribution. For very small $\gamma$, before entering the asymptotic regime 
the even $L^2$-distance $\Delta^2_+$ can become so small, that in experiments the cross over may not be observable at all. While non-spectral relaxation is a transient 
phenomenon in the Gaussian OUP for broad $\alpha$-stable initial distributions, the eigenvalues $\lambda_{ln}=-\nu(l+n\mu)$, $l,n\in\mathbb{N}$, used in the expansion 
of the conditional probability density, with a delta distribution as initial condition, are non-spectral for any $\mu\ne 2$, i.e. for a generalized L\'evy OUP.
\\ %\\
In conclusion, we have shown that the spectrum of the Hermitian counterpart of a Fokker-Planck operator corresponding to a 
Gaussian diffusion process in a potential only determines the time evolution of initial probability densities possessing all moments. 
Even then, broad initial distributions may relax slower than expected from the Hermitian eigenvalue spectrum, during a possibly long transient. 
This effect is a quite general property of relaxation from non-equilibrium initial conditions and must be taken into account in the interpretation of data in non-equilibrium systems. 
Furthermore, we have shown that the fractional Fokker-Planck operator for a L\'evy 
flight in a harmonic potential is, by similarity transformation, related to the Hamiltonian of a quantum harmonic oscillator. 
However, in this case, even a $\delta$-distribution, that is not located at the origin, can not be expanded into the transformed eigenfunctions 
of that Hamiltonian. 
Experimentally accessible quantities such as transition probabilities and autocorrelation functions of observables possessing second 
moments will therefore not relax at harmonic rates. While Hermitian operator spectral theory is a powerful tool to analyze a system, 
it is important to understand the limitations in its theoretical and experimental applications.

\end{document}